\documentclass[12pt]{article}
\usepackage{graphicx}

\newcommand\pubnumber{DPF2015-243}
\newcommand\pubdate{\today}
\newcommand{\pion}{\pi^{0}}
\def\napoli{University of Michigan\\
Ann Arbor, MI 48109}

\def\Title#1{\begin{center} {\Large #1 } \end{center}}
\def\Author#1{\begin{center}{ \sc #1} \end{center}}
\def\Address#1{\begin{center}{ \it #1} \end{center}}

\newcommand\pubblock{\rightline{\begin{tabular}{l} \pubnumber\\
         \pubdate  \end{tabular}}}
\newenvironment{Abstract}{\begin{quotation}  }{\end{quotation}}
\newenvironment{Presented}{\begin{quotation} \begin{center} 
             PRESENTED AT\end{center}\bigskip 
      \begin{center}\begin{large}}{\end{large}\end{center} \end{quotation}}

\def\beq{\begin{equation}}
\def\eeq#1{\label{#1}\end{equation}}
\def\eeqn{\end{equation}}

\def\beqa{\begin{eqnarray}}
\def\eeqa#1{\label{#1}\end{eqnarray}}
\def\eeqan{\end{eqnarray}}

\let\bar=\overbar

\def\Dslash{\not{\hbox{\kern-4pt $D$}}}
\def\dslash{\not{\hbox{\kern-2pt $\del$}}}

\def\msb{{\bar{\ssstyle M \kern -1pt S}}}

\begin{document}
\begin{titlepage}
\pubblock

\vfill
\Title{Parton Dynamics at PHENIX}
\vfill
\Author{ Joe Osborn}
\Address{\napoli}
\vfill
\begin{Abstract}
Investigating partonic interactions is one of the primary goals of the PHENIX experiment at the Relativistic Heavy Ion Collider (RHIC). RHIC is specially tailored for studying intrinsic partonic spin-momentum correlations due to its unique ability to collide polarized proton beams. Transverse single-spin asymmetries of order 10\% have been measured in PHENIX at center of mass energies from 62.4 GeV to 200 GeV, similar to previous measurements. These results indicate that there exist partonic transverse momentum effects within the proton and/or within the process of hadronization. The MPC-EX, a new silicon-tungsten preshower detector at PHENIX, has taken data for the first time this year with the intent of shedding further light on the origins of these asymmetries. A review of the status of the detector and of future planned measurements will be presented. An overview of ongoing work by PHENIX aimed at measuring intrinsic partonic transverse momentum will be discussed.
\end{Abstract}
\vfill
\begin{Presented}
DPF 2015\\
The Meeting of the American Physical Society\\
Division of Particles and Fields\\
Ann Arbor, Michigan, August 4--8, 2015\\
\end{Presented}
\vfill
\end{titlepage}
\def\thefootnote{\fnsymbol{footnote}}
\setcounter{footnote}{0}

\section{Introduction}

\begin{figure}[htb]
	\centering
	\includegraphics[scale=0.5]{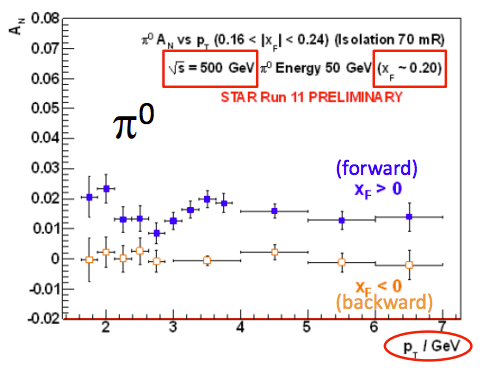}
	\caption{The STAR collaboration has measured the transverse single spin asymmetry for $\pi^0$s at a center of mass energy $\sqrt{s}=500$ GeV. The asymmetry is non-zero up to $p_T\approx 7$ GeV/c in the forward direction of the polarized proton, at $x_F\approx 0.2$. The asymmetry has been measured up to nearly 40\% from other collaborations at high $x_F\approx 0.8$ at a large range of center of mass energies \cite{anl,bnl,fnal,rhic}.}
	\label{fig:starpi0an}
\end{figure}

In transversely polarized proton-proton collisions, the transverse single spin asymmetry is an optimal observable to probe for partonic dynamics. The asymmetry is defined as follows:
\begin{equation}
A_N = \frac{\sigma^\uparrow(\phi)-\sigma^\downarrow(\phi)}{\sigma^\uparrow(\phi)+\sigma^\downarrow(\phi)}
\end{equation}
In the forward direction of the polarized proton, asymmetries that are orders of magnitude larger than the perturbative QCD (pQCD) prediction have been experimentally observed for a number of decades at energies well into the pQCD regime \cite{star,phenix}. Figure \ref{fig:starpi0an} shows an example of the $\pion$ asymmetry measured by the STAR collaboration, where the asymmetry is seen to be nonzero up to $Q^2\approx 49$ $[GeV/c]^2$ at small $x_F=\frac{2p_L}{\sqrt{s}}\approx 0.2$. At the center of mass energies that RHIC operates ($\sqrt{s}$=62.4-510 GeV), the PHENIX detector has an optimal opportunity to study transverse single spin asymmetries in a regime where pQCD is applicable yet these non-perturbative effects are still observable. In both experiment and theory we now have the tools to understand the origins of the asymmetries in terms of partonic dynamics since we have facilities that can observe the non-perturbative effects in a regime where theorists can interpret them with pQCD. \par

In order to theoretically understand these non-perturbative effects, Dennis Sivers introduced the idea of a Transverse Momentum Dependent Parton Distribution Function (TMD PDF)\cite{dennissiverspaper}. Transverse momentum dependent PDFs are written as a function of both x and the transverse momentum of the interacting partons $k_\perp^2$. Similarly, TMD fragmentation functions can also be written as a function of both z and $k_\perp^2$. Taking into account both the spin of the nucleon and the interacting quarks, there are several TMD PDF (and TMD FF) that quantify spin-momentum and spin-spin correlations between the nucleon spin and partonic transverse momentum (or spin). As an example, the Sivers TMD PDF quantifies a correlation between the initial-state proton spin and quark orbital angular momentum. On the other side of the hard scattering, the Collins TMD FF quantifies a correlation between the final-state quark spin and hadron orbital angular momentum. \par

Understanding these non-perturbative dynamics can be difficult in proton-proton collisions because of possible contributions from both the initial and final states. Separating contributions from initial-state effects and final-state effects is inconvenient and problematic because they are mixed up in hadron production. Direct photons, defined as photons coming directly from the hard scattering, are the optimal observable to ameliorate this problem. In $qg\rightarrow q\gamma$ and $q\bar{q}\rightarrow g\gamma$ the photon emitted does not suffer from hadronization or fragmentation effects, therefore it contains partonic information directly from the hard scattering. Understanding non-perturbative intitial-state effects with direct photons is one of the goals of the PHENIX experiment and the new MPC-EX detector recently installed in PHENIX. \par

\section{Experiment}
The PHENIX detector at the Relativistic Heavy Ion Collider (RHIC) is well suited to study these non-perturbative initial-state effects. RHIC is a versatile accelerator that can collide protons as well as many different species of heavy ions, such as Au, Al, and U. RHIC is also unique due to its ability to polarize one or both of the proton beams. Because of this RHIC is the only collider based facility in the world where studies of spin-momentum correlations are possible. In 2015, for the first time, RHIC also collided polarized proton beams on nuclei. This unique accelerator facility makes it possible to investigate spin-momentum correlations in many different species of collisions. \par

\begin{figure}[htb]
\centering
\includegraphics[scale=0.28]{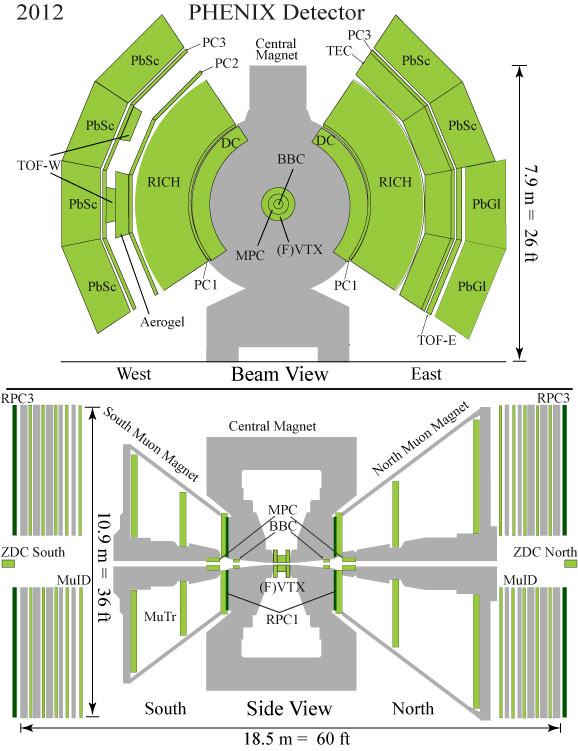}
\caption{The PHENIX detector configuration, shown both along the beam pipe (top) as well as from the side with the beam pipe going across the page from left to right (bottom).}
\label{fig:phenix}
\end{figure}

PHENIX is an optimal detector to search for non-perturbative initial-state effects due to its electromagnetic calorimetry at both midrapidity and forward rapidity. At midrapidity ($|\eta|<0.35$), PHENIX has two arms that each span $\Delta\phi=\pi/2$ radians with the capability for photon detection with an electromagnetic calorimeter as well as tracking of charged hadrons with a drift chamber, shown in the top of figure \ref{fig:phenix}. The EMCal has 8 sectors, 6 that are lead plastic scintillating towers, and 2 that are lead glass cherenkov detectors. The drift chamber allows for unidentified charged hadron tracking and has a 150 $\mu$m spatial resolution. The combination of the the EMCal and drift chamber allows for the analysis of a two-particle angular correlation measurement between direct photons and charged hadrons, which will be discussed further in section 3. \par

PHENIX also has electromagnetic calorimetry in the forward directions, at $3.1<|\eta|<3.8$, shown in the bottom of figure \ref{fig:phenix}. The Muon Piston Calorimeter, (MPC, named for its location in between the PHENIX muon magnets), is a lead tungstate calorimeter that was constructed in a region where spin-momentum correlations are known to be large. The MPC-EX (MPC-EXtension) is a new preshower detector that was installed at the end of 2014 to enhance the ability of the MPC. On its own, the MPC can only resolve $\pion\rightarrow\gamma\gamma$ decays up to $p^{\pion}=20$ GeV; at higher momenta the two photons merge into one cluster and cannot be resolved. The MPC-EX is a silicon tungsten preshower detector that will extend the momentum resolution of neutral pions in the forward region covered by the MPC up to 80 GeV. This increased resolution to $\pion$ decays will allow for measurements of $\pion$ asymmetries up to larger $x_F$, as well as larger $\pion$ background rejection for identification of forward direct photons.

\section{Initial-State Non-Perturbative Physics}

\begin{figure}[tbh]
\centering
\includegraphics[scale=0.6]{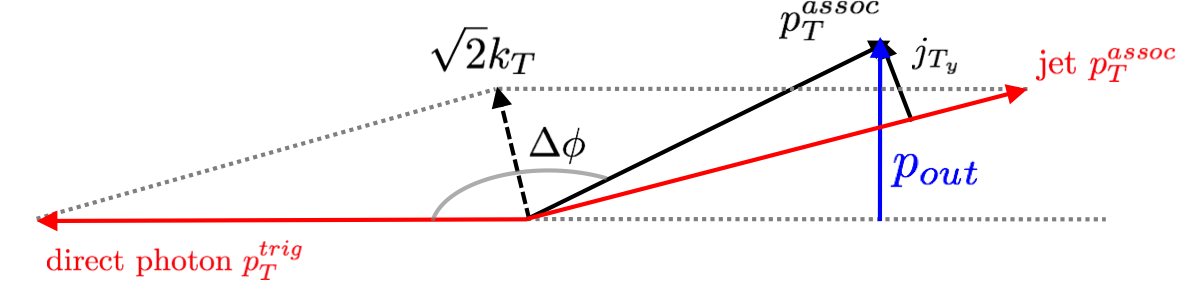}
\caption{Direct photon and away side charged hadron angular correlation kinematic diagram.}
\label{fig:dpkinematics}
\end{figure}

The intrinsic partonic transverse momentum is one fundamental quantity that characterizes the initial-state of partonic dynamics. One way to measure this quantity is through two-particle angular correlations, for example between a direct photon and away side charged hadron. Figure \ref{fig:dpkinematics} shows the vector kinematics of a direct photon and away side charged hadron, where the direct photon is represented by the red vector on the left and the away side hadron fragments from the recoiled quark or gluon on the away side, shown by the black vector on the right of the nominal interaction point. The intrinsic transverse momentum can be seen from this vector diagram, and quantifies the vector sum of the two initial partons intrinsic transverse momentum. 

\begin{figure}[tbh]
\centering
\includegraphics[scale=0.28]{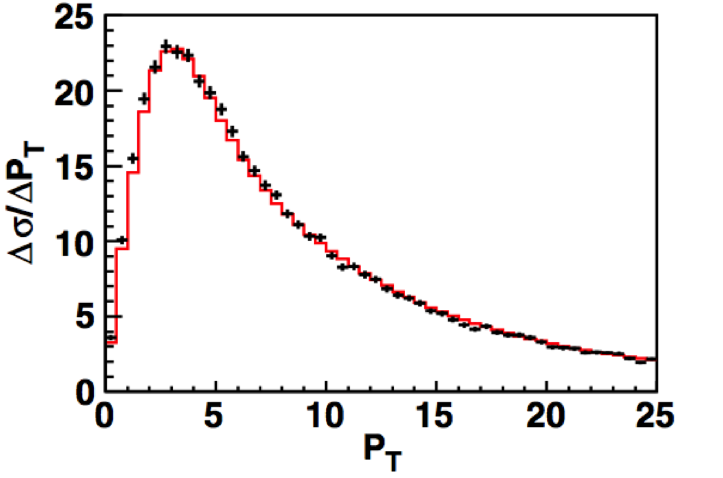}
\caption{Drell Yan Z Boson production cross section from CDF \cite{cdfzboson}.}
\label{fig:zcross}
\end{figure}

The effects of intrinsic transverse momentum can be seen in distributions as simple as a cross section, such as the CDF Drell-Yan Z boson cross section. For Drell-Yan Z boson production at small $p_T$, the $p_T$ of the Z boson will be generated solely by the intrinsic partonic transverse momentum of the $q\bar{q}$ pair that annihilated. At large $p_T$ figure \ref{fig:zcross} shows that the cross section exhibits a power law shape which is described by pQCD. At small $p_T$, where the transverse momentum is generated by intrinsic $k_T$, the cross section turns over and exhibits a gaussian like structure. This transition from pQCD power law to gaussian shape is characteristic of where non-perturbative effects become evident. In the case of angular correlations, similar behavior is expected from the vector $p_{out}$ in figure \ref{fig:dpkinematics}. This vector characterizes the out of jet plane transverse momentum of the fragmented hadron, and at small $p_{out}$ is generated mainly from the intrinsic transverse momentum of the colliding partons. Therefore one would expect similar behavior in the $p_{out}$ distribution from angular correlations to the CDF Z boson cross section plot at small $p_T$. This will help characterize the intrinsic transverse momentum of the colliding partons. At PHENIX, there is an ongoing two particle correlations analysis in the high statistics Run 13 $\sqrt{s}=510$ GeV $p+p$ data set that will investigate these initial-state non-perturbative effects due to intrinsic transverse momentum. \par

In the forward direction, investigations of non-perturbative initial-state physics will be greatly benefited by the addition of the MPC-EX. Because of the increased resolution to $\pion$ production via their 2 photon decay in the forward direction, measurements of the transverse single spin asymmetry for $\pion$s will be possible up to large $x_F\approx 0.8$ at $\sqrt{s}$=200 GeV. More importantly the increased resolution to $\pion$ production will also allow for increased decay photon rejection, and thus an increased signal to background ratio for direct photons in the forward direction. Measuring the direct photon single spin asymmetry is an important measurement for the purpose of understanding what initial state effects are generating the large asymmetries observed. Understanding if the asymmetries are generated by a Sivers type effect or Collins type effect (or both, and by how much) would be an important step to learning what underlying processes could be generating the large transverse single spin asymmetries that have been observed in many different types of collisions and center of mass energies. In 2015 the MPC-EX finished taking data for the first time, collecting an integrated luminosity of approximately 60 pb$^{-1}$ of $p^\uparrow+p$, 205 nb$^{-1}$ of $p^\uparrow+Au$, and 450 nb$^{-1}$ of $p^\uparrow+Al$ collisions. Analysis efforts are currently ongoing and should produce important and interesting physics results that help unveil how initial-state non-perturbative effects are generating the large asymmetries observed.

\end{document}